# Short Gamma-Ray Bursts Are Different

J.P. Norris[1], J.D. Scargle[2], and J.T. Bonnell[1]

[1] *NASA/Goddard Space Flight Center, Greenbelt, MD 20771, USA*
[2] *NASA/Ames Research Center, Moffett Field, CA 94035-1000, USA*

**Abstract.** We analyze BATSE time-tagged event (TTE) data for short gamma-ray bursts ($T_{90}$ duration < 2.6 s), studying spectral lag vs. peak flux and duration, as well as the number of distinct pulse structures per burst. Performing the cross-correlation between two energy bands, we measure an average lag ~ 20–40 × shorter than for long bursts, and a lag distribution close to symmetric about zero – unlike long bursts. Using a "Bayesian Block" method to identify significantly distinct pulse peaks, we find an order of magnitude fewer pulses than found in studies of long bursts. The disparity in lag magnitude is discontinuous across the ~ 2-s valley between long and short bursts. Thus, short bursts do not appear to be representable as a continuation of long bursts' temporal characteristics.

## 1   Introduction

Our understanding of short bursts is limited to their gamma-ray characteristics, whereas for long bursts we have multi-wavelength afterglows, redshifts, host galaxies, evidence of progenitors arising in star-forming regions, plus lots of theory. For short bursts, we have the "Smaller Bump" in the GRB duration distribution. The following references serve as a summary of the sparse research history of short bursts: their identification as a subclass by duration [5]; early indications that their fraction is ~ 1/4 of the total [6]; their harder spectra [3]; and truncation effects that could be hiding even more bursts at shorter durations [13]. Most relevant to the present work, was the suggestion by two groups that a continuous deformation of some combination of long bursts' temporal characteristics – the distributions of number of pulses per burst, pulse width, and intervals between pulses – could explain short bursts' durations and the 2-s minimum [7,16]. Here, we further quantify the temporal differences between long and short GRBs, finding the properties of the two classes to be disjoint.

## 2   Spectral Lag and Pulse Analysis

The sample comprises all 261 short bursts (durations measured via technique similar to that in reference [1]) with "good" TTE data (the burst was contained within the TTE 32k event buffer), and peak flux (PF) > 2 photons cm$^{-2}$ s$^{-1}$ (50–300 keV). We used a "Bayesian Block" cell coalescence method [14,15] to determine the burst region and significant peaks and valleys, yielding the number of pulses per burst. We cross-correlated the 25–50 keV and 100–300 keV energy channels, bootstrapping the TTE data to make 51 profiles per burst. We fitted a cubic near the peak of the cross correlation function (CCF), requiring 51 consecutive fits to be concave down; else, we lengthened the fitted region, repeating the procedure for a new set of 51 profiles. The mode of the CCF peaks per burst was taken as the spectral lag measure, and error bars were generated.

## 3   Lag Results for Short Bursts

Whereas CCF lags for long bursts [9,10] extend up to ~ 2 s, with the core near 50 ms, the maximum lags for short bursts are ~ 30 ms, or ~ 60 × shorter (Figure 1a). Virtually all long bursts analyzed (save for two apparently anomalous cases) have positive lags, or insignificantly negative determinations [9,10]. Figure 1a shows that short burst lags are distributed approximately symmetrically about zero lag with the mode near 0{2 ms. The better determined lags at higher peak flux are also concentrated near zero. In fact, for PF > 10 photons cm$^{-2}$ s$^{-1}$, the absolute value of the CCF lag is less than 2.5 ms for 25 of 35 bursts. When divided into four duration ranges as shown in Figure 1b, short bursts do not evidence a trend towards longer lags for longer short bursts: The core of the lag distribution is contained within ± 15 ms with modes of less than ± 5 ms, essentially independent of duration. Similarly, a duration-independent discontinuity in lag is evident for long bursts as well [11].

From our Bayesian Block coalescence analysis, we determine that bright short bursts usually have ~ 1–2 pulses per burst [11]. In comparison, bright long bursts have a broad range in number of pulses – a few to tens – as can be appreciated from Figure 5 of Lee et al. [4] (see also [8]). Fewer pulses are discernible in weaker short bursts [11], a brightness bias also evident in long bursts [4]. Thus long bursts average an order of magnitude more pulses per burst than do short bursts. Combined with the discontinuity in lag, the results suggest that the temporal properties of the two classes are disjoint, in the vicinity of ~ 2 s.

## 4   Summary and Conclusions

Short bursts appear to be a "different" phenomenon (by Harwit criteria [2]): Their lags are of order ~ 20–40 × shorter than lags for long bursts, with the discontinuity near ~ 2 seconds ($T_{90}$ duration). The lag distribution appears close to symmetric, about zero. They have an order of magnitude fewer pulses than do long bursts. In the duration distribution the two modes are separated by a factor of order 100, ~ 250 ms compared to ~ 25 s. Strengthening this picture is the analysis of Paciesas et al. [12] which finds a discontinuity in the distribution of $E_{peak}$ values between short and long bursts.

Thus, based on our knowledge of their gamma-ray characteristics alone, we conclude that short bursts cannot be represented as a continuation of long bursts. The two classes are disjoint, temporally and spectrally. Moreover, short bursts, by their nature, continue to be a difficult study in terms of their physical properties: They appear to be the only remaining high-energy astrophysical phenomenon with no Rosetta Stone at longer wavelengths.

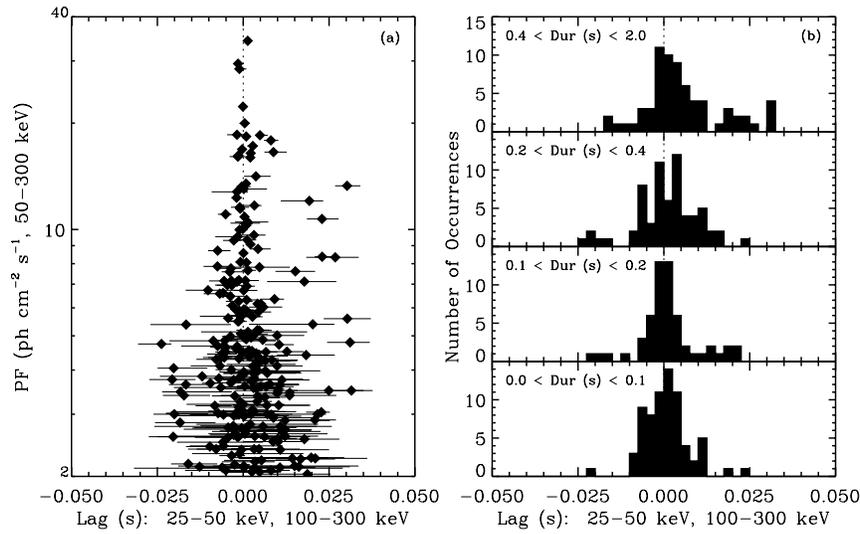

**Fig. 1.** (a) CCF lag vs. peak flux for short bursts. (b) Histograms of CCF lags for four duration ranges, illustrating that the mode is nearly independent of duration.